\documentclass[10pt,a4paper,twoside]{article}
\usepackage[T1]{fontenc}
\usepackage[utf8]{inputenc}

\usepackage{amssymb}
\usepackage{amsmath}
\usepackage{physics}
\usepackage{qcircuit}
\usepackage{enumerate}
\usepackage{booktabs}

\usepackage{fancyhdr}
\usepackage{verbatim}
\usepackage{lastpage}
\usepackage{ifthen}

\usepackage[pdftex]{color,graphicx}
\usepackage{hyperref}
\usepackage{iberamia}   

\newcommand{\card}[1]{\ensuremath{\left\|#1\right\|}}

\newcommand{\thispaperdoi}{}
\title{Quantum circuit synthesis of Bell and GHZ states using projective simulation in the NISQ era}

\author{Otto Menegasso Pires$^{[1]}$, Eduardo Inacio Duzzioni$^{[2]}$, Jerusa Marchi$^{[1]}$, Rafael de Santiago$^{[1]}$, 
\smallskip \\
\small{$^{[1]}$INE - Departamento de Informática e Estatística, at Universidade Federal de Santa Catarina, Florianópolis, Brazil
\smallskip \\
$^{[2]}$FSC - Departamento de Física, at Universidade Federal de Santa Catarina, Florianópolis, Brazil
}}
\date{}

\begin{document}

\medskip
\maketitle

\pagestyle{Rest}
\thispagestyle{FirstPage}
\smallskip

\noindent{\small{{\textbf Abstract}
Quantum Computing has been evolving in the last years. Although nowadays quantum algorithms performance has shown superior to their classical counterparts, quantum decoherence and additional auxiliary qubits needed for error tolerance routines have been huge barriers for quantum algorithms efficient use.
These restrictions lead us to search for ways to minimize algorithms costs, i.e the number of quantum logical gates and the depth of the circuit. For this, quantum circuit synthesis and quantum circuit optimization techniques are explored.
We studied the viability of using  Projective Simulation, a reinforcement learning technique, to tackle the problem of quantum circuit synthesis for noise quantum computers with limited number of qubits. The agent had the task of creating quantum circuits up to 5 qubits to generate GHZ states in the IBM Tenerife (IBM QX4) quantum processor.
Our simulations demonstrated that the agent had a good performance but its capacity for learning new circuits decreased as the number of qubits increased.}}

\medskip



\section{Introduction}
\label{sec:introduction}
    Quantum computing is a promising new paradigm for computer science. Quantum algorithms have proven themselves superior to classical ones in some classes of problems. The major examples are Shor’s algorithm for solving the hidden subgroup problem and Grover’s algorithm for the unstructured search problem, but quantum computing is not limited to them\cite{Montanaro:2016}. Simulating complex atomic systems using a quantum computer\cite{quantum-simulation} could profoundly impact physics research since a quantum computer can naturally simulate the effects of quantum mechanics.
    
    Quantum Computing has also been impactful in the machine learning field. Basic linear algebra subroutines, such as Fourier transform, finding eigenvectors and eigenvalues, and solving linear equations, for example, exhibit exponential quantum speedups over their best known classical counterparts\cite{QML} Also, quantum machine learning aims to implement machine learning algorithms in quantum systems, by using the quantum properties such as superposition and entanglement to solve these problems efficiently\cite{QML_review}.
    
    However, practical limitations still hinder the development of a universal quantum computer that could use such algorithms.
    Current quantum computers are referred to as NISQ (Noisy Intermediate-Scale Quantum) computers \cite{NISQ} because of qubits imperfections and their available limited number, currently between 50 and 100. These numbers are not sufficient to execute error-correcting codes once 9 qubits are necessary to make 1 qubit fault-tolerant \cite{Devitt_2013}. Another limiting factor of NISQ computers is qubit connectivity. Not all qubits are directly connected, which limits the available two-qubit gate operations. This problem can be solved by the insertion of SWAP gates to change the qubit mapping\cite{Nisq-map}, but due to imperfections in NISQ quantum gates, the remap of qubits introduces more errors into the system. 
    
    While we do not have the necessary resources to make an efficient quantum computer, an alternative solution is to make the best of the currently available resources through the development of techniques for quantum circuit synthesis and quantum circuit optimization. This paper focuses on quantum circuit synthesis.
    
   Quantum circuit synthesis is mainly made through the synthesis of reversible boolean circuits\cite{AIG-MIG}\cite{ESOP}\cite{BDD}\cite{DFS-BFS}. These techniques focus on generating a classical reversible circuit and then matching it with an equivalent quantum circuit. This is possible because quantum computing is naturally reversible. Hence, it is easy to convert a reversible boolean circuit to a quantum circuit. 
    
    Techniques to synthesize quantum circuits directly such as genetic programming\cite{genetic09}\cite{GP-quantum} or swarm algorithm\cite{swarm} have also been proposed. They are an interesting alternative once it enables the optimization of the circuit over the synthesis process.
    
    Reinforcement learning also presented good results in a correlated problem. In  \cite{Zeilinger:2018} a reinforcement learning technique called Projective Simulation \cite{PS-original} was used to develop an artificial agent capable to discover new optical quantum experiments by learning new placements for the optical elements.
    
    In this sense, this paper proposes the application of the Projective Simulation reinforcement learning technique as a new method for quantum circuit synthesis. We demonstrate that reinforcement learning could be a valuable tool for quantum circuit synthesis: the agent could learn alternative circuit layouts to perform computations. However, we first present how to model the problem of quantum circuit synthesis in a way an artificial agent could learn it.
    
    The paper is structured as follows. Section \ref{sec:background} explains the basics of quantum computing and projective simulation. Section \ref{sec:Method} describes the agent's model and the simulation environment parameters, detailing our contributions. Section \ref{sec:results} shows the results obtained to different circuit configurations. In Section \ref{sec:conclusion} we discuss the results and propose future developments of this work.
    
\section{Background}
\label{sec:background}
This section presents a short introduction to quantum computing and quantum circuits. We focus on Bell states as the target circuits we want to synthesize. In Section \ref{ssec:Projective} we introduce the reinforcement learning technique used in our proposal.

\subsection{Quantum Computing and Quantum Circuits}
\label{ssec:quantumC}

A quantum bit (qubit) is the basic unit of quantum information. A qubit is represented by a unit vector in a Hilbert space. Let $\ket{0}$ and $\ket{1}$ the computational basis of quantum computing. They are represented by the vectors
\[
\ket{0} = \begin{bmatrix} 1 \\ 0 \end{bmatrix} \hspace{2em}
\ket{1} = \begin{bmatrix} 0 \\ 1 \end{bmatrix}.
\]

A qubit can take the value of the computational basis and also any intermediate value between $\ket{0}$ and $\ket{1}$. The qubit is then described as a linear combination of the vectors forming the basis:
\[
    \ket{\psi} = \alpha\ket{0} + \beta\ket{1}
\]
where $\alpha$ and $\beta$ are complex numbers. These intermediate states are called \textit{superpositions}. 

A multiple qubit system follows the notation \[\ket{00} = \ket{0}\ket{0} = \ket{0} \otimes \ket{0},\] where $\otimes$ is the tensor product operation. Tensor product is a way to put qubits together and form larger vector spaces.

When a system of two qubits cannot be described as the tensor product between the qubits that make it, the qubits are \textit{entangled} \cite{horodecki2009quantum}. Figure \ref{fig:Bell} describes the entangled states of a 2-qubit system, called Bell's states, and of a 3-qubit system, called $GHZ$-state \cite{greenberger1990bell,greenberger1989going}. Entanglement and superposition are properties explored by quantum algorithms to achieve speedup over classical ones.\cite[Chapter 1.3.7, 1.4]{Nielsen:2010}

\begin{figure}[h]
    \[ \frac{\ket{00} + \ket{11}}{\sqrt{2}}   = \ket{\beta_{00}}, \]
    \[ \frac{\ket{01} + \ket{10}}{\sqrt{2}}   = \ket{\beta_{01}}, \]
    \[ \frac{\ket{00} - \ket{11}}{\sqrt{2}}   = \ket{\beta_{10}}, \]
    \[ \frac{\ket{01} - \ket{10}}{\sqrt{2}}   = \ket{\beta_{11}}, \]
    \[ \frac{\ket{000} + \ket{111}}{\sqrt{2}} = GHZ\text{-state}. \]
    \caption{A set of entangled states: Bell's states are obtained by the entanglement of 2 qubits and $GHZ$-state is obtained by 3 qubits.}
    \label{fig:Bell}
\end{figure}

Operations in quantum computing are described by $2^n \times 2^n$ dimension unitary complex-valued matrices where $n$ denotes the number of qubits in the quantum system. A unitary matrix follows the rule $U^{\dagger}U = UU^{\dagger} = I$, whereas $U^\dagger$ is the Hermitian conjugate of U and I is the identity matrix. These operations are known as quantum logic gates. Figure \ref{fig:pauli} shows some examples of common single-qubit gates.
Gates $X$, $Y$, and $Z$ rotate the qubit around the axis $x$, $y$, and $z$, respectively and are known as Pauli Matrices. The $X$ gate is also known as the quantum NOT gate, because it sends the qubit from the state $\ket{0}$ to the state $\ket{1}$ and vice-versa. The $H$ gate is the Hadamard gate. It sends the qubit from the computational basis to a state ``half the way'' from $\ket{0}$ to $\ket{1}$. All these gates are single-qubit gates.
The $CNOT$ gate is the controlled-not gate.

    \begin{figure}[ht!]
        \centering
        $
        X = \begin{bmatrix} 0 & 1  \\  1 &  0 \end{bmatrix}
        \hspace{2em}
        Y = \begin{bmatrix} 0 & -i \\  i &  0 \end{bmatrix}
        \hspace{2em}
        Z = \begin{bmatrix} 1 & 0  \\  0 & -1 \end{bmatrix}
        $
        
        \bigskip
        $
        H = \begin{bmatrix} 1 & -1 \\  1 &  1 \end{bmatrix}
        \hspace{2em}
        CNOT = \begin{bmatrix} 1 & 0 & 0 & 0  \\ 0 & 1 & 0 & 0  \\ 0 & 0 & 0 & 1  \\ 0 & 0 & 1 & 0 \end{bmatrix}
        $
        
        \caption{Matrix representation of some basic quantum gates, notably Pauli (X,Y,Z), Hadamard and CNOT.}
        \label{fig:pauli}
    \end{figure}

All quantum logic gates can be transformed into a controlled-gate. Controlled gates have two sets of qubits, target qubits and control qubits. Once control qubits are in the correct value - $\ket{1}$ for positive controls or $\ket{0}$ for negative controls -  the gate is applied on the target qubits. When drawing a circuit, a control qubit is represented by a bullet connected by a line to the gate it is related to.

\begin{figure}[h!]
        \centerline{
        \Qcircuit @C=1em @R=0.7em {
            \lstick{\ket{0}} & \gate{H} & \qw      & \ctrl{1} & \qw \\
            \lstick{\ket{0}} & \gate{X} & \ctrl{1} & \gate{Z} & \qw \\
            \lstick{\ket{0}} & \qw      & \targ    & \qw      & \qw
            }
            \hspace{1cm}
        }
        \caption{A generic quantum circuit composed of two single qubit gates, a Hadamard (H) and a X gate, and two controlled gates, a CNOT and a controlled-Z gate.}
        \label{fig:circuit-example}    
\end{figure}
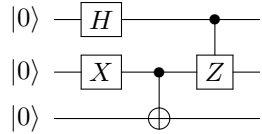

In a circuit, a qubit is represented by a horizontal line. Quantum logic gates are represented by labeled boxes, with few exceptions. The computation is performed from left to right, by the application of the gates' operation over the qubits. The maximum number of sequential gates is called the circuit depth. Figure \ref{fig:circuit-example} shows an example of a quantum circuit composed of 3 qubits and 4 quantum logic gates, 2 single-qubit gates, and 2 positive controlled-gates. 

The circuit starts by applying a Hadamard gate on the first qubit and an X gate on the second qubit. Then, there is a CNOT operation between the second and third qubits, written in its most common form. The last operation is a controlled-$Z$ operation between the first and the second qubit.

Current quantum circuits have physical problems that should be considered. Qubits have a maximum lifespan before they degenerate into noise, called \textit{decoherence time}, and quantum logic gates have an error rate caused by imperfections in implementation. When these limitations are considered, a reliable quantum circuit should have gates with minimum error and shorter depths.


\subsection{Projective Simulation}
\label{ssec:Projective}

    \textit{Projective Simulation} (PS) is a learning model based on stochastic processing of an episodic memory that can be applied in reinforcement learning solutions. This technique was introduced by \cite{PS-original} and was further developed by \cite{PS-definition}.

    The PS agent is an entity located in a partially unknown environment, that receives perceptions through its sensors. The agent acts in response to the environment and its actions are rewarded by the environment.
    
    The agent has a specific type of memory named Episodic and Compositional Memory (ECM). The ECM is represented by a weighted directed graph, where nodes are called \emph{clips}. Clips are the basic memory units corresponding to single episodic experiences, as perceptions (perception-clip), action (action-clip), or a combination of both (perception-action-clip). A clip can be excited when the agent receives the actual state of the environment. The excitation hops to adjacent clips with some probability related to the weight of the edge connecting those clips. Thus each coming perception from the environment starts a random walk through the ECM. The random walk ends when an action-clip is reached, which generates a real action in the environment.

    The learning process of PS is defined through modifications in the agent's ECM. These modifications can be the addition or the deletion of a clip or the edge weight update between clips. At the beginning of the simulation, the ECM is a \textit{tabula rasa}, i.e. the agent has no preferences towards a specific kind of action. According to the received reward from the environment, the agent's ECM is updated following a predefined set of rules. The changes in the agent's ECM result in a new decision making, that leads to a new reward from the environment, which changes the structure of the ECM. Since the PS agent has no explicit model of the environment which predicts the next state or reward, it is considered model-free.
    
    The agent has the probability $P^{(t)}(a|s)$ of making the action \textit{a} given the perception \textit{s}. However, a complete description of the agent connects the probability $P^{(t)}(a|s)$ with its memory internal state at time \textit{t}, and specifies how the memory is updated as the agent interacts with the environment.

    In order to clarify the main concepts of the PS model, we list some formal terms used in PS:

        \begin{itemize}
            \item \textbf{Episodic and Compositional Memory (ECM):} is the main element of PS, defined as a weighted directed graph. Each node is called a clip.
            \item \textbf{Clip:} represents short episodic experiences. Clips are defined by L-tuples $c = (c_1, c_2, ..., c_L)$. Each $c_i$ is either a perception ($s \in S$) or an action ($a \in A$), where both are defined below.
            \item \textbf{Perceptions:} are defined as $N$-tuples $s = (s_1, s_2,...,s_N) \in S \equiv S_1 \times S_2 \times ...  \times S_N, \> s_i \in {1,...,\card{S}}$, where the number of possible perceptions is given by $S \equiv \card{S} \equiv \card{S_1}...\card{S_N}$.
            \item \textbf{Actions:} are defined as $M$-tuples $a = (a_1, a_2,...,a_M) \in A \equiv A_1 \times A_2 \times ... \times A_M, \> a_i \in {1,...,\card{A}}$, where the number of possible actions is given by $A \equiv \card{A} \equiv \card{A_1}...\card{A_M}$.
            \item \textbf{Edges:} Each edge connecting clip $c_i$ to clip $c_j$ has a dynamic weigth $h^{(t)}(c_i,c_j)$, called  $h$-value, which changes over time.
            \item \textbf{Hopping Probability:} The probability of an excitation hopping from clip $c_i$ to clip $c_j$ is given by \[p^{(t)}(c_i,c_j) = \frac{h^{(t)}(c_i,c_j)}{\sum_k h^{(t)}(c_i,c_k)}\] which represents the sum of the $h$-values of all edges connecting the clips $c_k$ to $c_i$  for all clip $k$.
            \item \textbf{Interface:} The interface between the PS agent and the environment is made with its sensors and actuators and their connection to memory. An external perception $s$ excites a certain perception-clip $c$ according to the function $\mathcal{I}(c|s)$. Similarly, an action-clip couples out to perform a real action $a$ according to the function $\mathcal{O}(a|c)$. The input function $\mathcal{I}(c|s)$ and the output function $\mathcal{O}(a|c)$ connect the internal random walk, described by the hopping probability $p^{(t)}(c_j|c_i)$, with the external behaviour of the agent, described by $P^{(t)}(a|s)$.
            \item \textbf{Damping Parameter:} PS admits an optional parameter $\gamma$ called damping parameter, usually $(0\leq\gamma\leq 1)$. This parameter can be seen as the agent ``forgetting'' over time. This value weakens the connections between clips, allowing the agent to adapt to changing environments.
            
        \end{itemize}

        The process behind PS is stochastic. Each step $t$ begins with a perception from the environment exciting a memory clip $c_i$ according to function $\mathcal{I}(c|s)$. Then, the excitation hops from clip $c_i$ to an adjacent clip $c_j$, with the probability $p^{(t)}(c_j|c_i)$. This process continues as a random walk, allowing the excitation to move through the clip network. The walk ends when an action-clip is reached and generates an equivalent action in the environment according to function $\mathcal{O}(a|c)$. Each action receives from the environment a reward $\lambda \ge 0$ that is used, together with damping parameter $\gamma$, to update all the $h$-values as follows: 
        
        \begin{enumerate}[(i)] 
            \item For activated edges, i.e. the edges that were crossed in the last random walk
        \[
        h^{(t+1)}(c_i,c_j) = h^{(t)}(c_i,c_j) - \gamma (h^{(t)}(c_i,c_j) -1) + \lambda
        \]
        where $t$ is the current step, $\gamma$ $(0\leq\gamma\leq 1)$ is the damping parameter and $\lambda$ is the reward given by the environment.
        
        \item For other edges the $h$-values are damped
        \[
        h^{(t+1)}(c_i,c_j) = h^{(t)}(c_i,c_j) - \gamma (h^{(t)}(c_i,c_j) -1)
        \]
        \end{enumerate}
        
 
    When the agent earns a positive reward, the edges that were crossed during the random walk that ended in the correct action are strengthened. These edges then have a greater probability of being chosen again in future steps. Still, if a wrong action is taken and no reward is earned, i.e. $\lambda=0$, all edges are damped, including those that were crossed in the last random walk. 


    \begin{subsubsection}{Temporal Correlation}
    
        In some reinforcement learning problems, the reward is not related to just the last action made by the agent, but also to previous actions. This case is referred to as Temporal Correlation.
    
        PS allows the use of temporal correlation through a new parameter in the reward function. This parameter is called edge glow and represents the level of excitement of an edge. For example, every time an edge is chosen, the agent sets the edge glow to the maximum value, and for each step that it is not chosen its glow fades. Thus an edge that was not chosen in the last random walk can still be partially rewarded. This effect is referred to as \emph{afterglow}.
        
        Formally, the afterglow is implemented by adding a new parameter $g$ called glow, which is related to the ECM's edges. Initially, $g=0$ for all edges. So, anytime an edge is crossed in the random walk it sets $g=1$. Every subsequent step $g$ decays by a rate $\eta$ until reaches 0, following the rule:
        \[
            g^{(t+1)} = g^{(t)} - \eta g^{(t)}, 0 \le \eta \le 1,
        \]
        and the reward function chages to:
    \[
        \begin{array}{lll}
        h^{(t+1)}(c_1,c_2) = & h^{(t)}(c_1,c_2) - \gamma(h^{(t)}(c_1,c_2)-1)\\
        & + \lambda g^{(t)}(c_1,c_2)
    \end{array}
    \]
    
    \noindent where $g^{(t)}(c_1,c_2)$ represents the value $g$ of the edge connecting $c_1$ and $c_2$ at time $t$. As before, only the excited edges are strengthened by the reward $\lambda$. The difference is that the edges that were partially excited can also be partially rewarded.

    \end{subsubsection}

    \begin{subsubsection}{Composition}
        
        The compositional aspect of the ECM is a dynamic process that allows structural changes in its inner connections. In particular, it allows the spontaneous creation of new clips inside the ECM, through the combination or variation of existing clips. The new clips can represent fictitious episodes that were not tried before, thus expanding the diversity of conceivable events and actions that exists inside the ECM. As a result, the memory is less connected with the agent's real past. This effectively enables the agent to create alternative options that it has already found previously, thus making it more capable and flexible.
    
        There are many possibilities to merge clips. In \cite{PS-definition} a merging mechanism is defined for $M$-dimension action clips, considering the following cases:

        \begin{itemize}
            \item Two action clips $c_a = (a_1, a_2, ..., a_M)$ and  $c_b = (b_1, b_2, ..., b_M)$ are merged in a new clip if and only if: 
            \begin{enumerate}[(a)]
            \item Both corresponding actions were sufficiently rewarded through the same perception; 
            \item The action clips $c_a$ and $c_b$ differ exactly in two components.
            \end{enumerate}
            
            \item When action clips $c_a$ and $c_b$ differ in theirs $i$-th and $j$-th components, the merge results in two new composed clips:
            $$c^{new}_1 = (a_1, a_2, ..., b_i, ..., a_j, ..., a_M)$$ 
            
            \noindent and 
            
            $$c^{new}_2 = (a_1, a_2, ..., a_i, ..., b_j, ..., a_M)$$
            
            \item A new clip is only created if it is not already in the agent's ECM.

            \item New action clips are connected to corresponding perception clip $c_0$ with $h$-value equals to the sum of the $h$-values of the originals action clips. 
            
            $$h(c_0, c^{new}_1) = h(c_0, c^{new}_2) = h(c_0, c_a) + h(c_0, c_b)$$
            
            \noindent Furthermore, new action clips are connected to all other perception clips with initial $h$-value equals to $1$.

        \end{itemize}
   
    \end{subsubsection}
\section{Synthesizing Quantum Circuits with Projective Simulation}
\label{sec:Method}

    Inspiring by the good results obtained by \cite{Zeilinger:2018} in optical quantum experiments, we used Projective Simulation to develop a quantum circuits synthesizer. Our first intent was to explore the capability of reinforcement learning to synthesize circuits capable of generating entangled states. We use PS with the optional damping parameter $\gamma$ as well as temporal correlation.
    
    We defined the environment as a quantum circuit, following the architecture Tenerife from IBM Quantum Experience \footnote{https://quantum-computing.ibm.com/}. The circuit has a predetermined number of qubits (varying from $2$ until $5$) and at the beginning of the learning process, it does not have any gate. 
    
    The agent is described by its episodic and compositional memory (ECM) whose initial perception-clip is the tensor product of the qubits, i.e, the environment's initial state. We set the action pool as gates and the agent's action-clip is the application of one gate in the circuit, i.e. the agent can choose an action from the set of gates composed by Hadamard, $X$, $Y$, $Z$ and $CNOT$ gates. 
    
    The agent can place the gate at any qubit in the circuit, always after all the other gates already placed. 
    We chose this architecture to limit the possibilities of placement for the CNOT gate.
    
    Every time the agent places a new gate in the circuit the next state is generated by multiplying the gate unitary matrix to the tensor product of the qubits. The agent receives the qubits as a new perception and looks for a corresponding perception-clip. If none is found a new clip is created that matches with the perception.
    
    For practical reasons, if the episode ends without a reward, the agent deletes every recently created perception-clip in the episode. Otherwise, the consumption of memory would spike fastly. 
    
    We chose to apply the following reward function:
    
    \begin{equation}
    \lambda = base\_value - \sum e_g * \frac{d_{min}}{d_i}    \end{equation} 
    
    \noindent where $e_g$ is the error of each gate put in the circuit, $d_{min}$ is the depth of the minimum circuit found, $d_i$ is the depth of the circuit being rewarded and $base\_value$ is a predetermined fixed value that changes with the number of qubits. We set $d_{min}$ to starts equal to the maximum depth allowed by the circuit and it is updated dynamically during the learning process. The $base\_value$ was set to $100$ for $2$-qubits circuits and it is increased by $50$ for each new qubit, for a total of $250$ for $5$ qubits. The reward was based on the reward used by \cite{Zeilinger:2018} to learn new quantum experiments. They used a reward value $\lambda = 100$ without any penalties being calculated.
    
    We also added penalties in the reward function to make the agent consider physical limitations and prioritize better circuits. One circuit is better if the used gates have minimum error and if the depth is shorter. Thereby, the operation made by the circuit will be less affected by qubit decoherence time and errors resulted from technological implementations.
    
    We made 4 simulations, generating circuits from 2-qubits to 5-qubits. The used parameters of each simulation are summarized in Table \ref{tab:parameters}. 
    
    \begin{table*}[th!]
        \centering
        \begin{tabular}{cccccc}
        \toprule
        Parameter & Simulation 1 & Simulation 2 & Simulation 3 & Simulation 4  \\
        \midrule
            Number of Qubits               & 2   & 3   & 4   & 5   \\
            Damping Parameter $\gamma$     & 0.1 & 0.1 & 0.1 & 0.1 \\
            Parameter $\eta$ (afterglow)   & 0.1 & 0.1 & 0.1 & 0.1 \\
            Maximum Circuit Depth          & 4   & 5   & 6   & 7   \\
            Base Value of Reward $\lambda$ & 100 & 150 & 200 & 250 \\
        \bottomrule
        \end{tabular}        \caption{Tunning for each simulation}
        \label{tab:parameters}
    \end{table*}

    We set the damping parameters $\gamma = 0.1$ and $\eta = 0.1$ for all performed simulations. We defined a maximum circuit depth for the circuits to limit the size of each episode. The maximum depth was 4 gates, for circuits with 2-qubits, 5 gates for circuits with 3-qubits, 6 gates for circuits with 4-qubits, and 7 gates for 5-qubits. The values were chosen based on the goal of the simulation, described below. 
        As allowing greater depths would not make it easier for the agent to learn, it could actually hinder its progress, we decided to add a margin of two gates for the agent to learn the right circuit.
    
        \begin{figure}[b]
            \centerline{
            \Qcircuit @C=1em @R=.7em
                { \lstick{\ket{0}} & \gate{H} & \ctrl{1} & \qw      & \qw & \multimeasureD{2}{\frac{\ket{000}+\ket{111}}{\sqrt{2}}} \\
                \lstick{\ket{0}} & \qw      & \targ    & \ctrl{1} & \qw & \ghost{\frac{\ket{000}+\ket{111}}{\sqrt{2}}} \\
                \lstick{\ket{0}} & \qw      & \qw      & \targ    & \qw & \ghost{\frac{\ket{000}+\ket{111}}{\sqrt{2}}} } }
            \caption{Basic circuit for generating a 3-qubit GHZ state.}
            \label{fig:ghz_circuit}
        \end{figure}
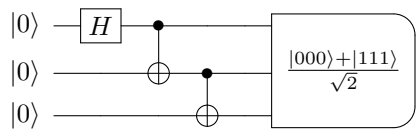
        
    The goal of each simulation was to find a circuit capable of generating entangled states. The goal states were the Bell state $\ket{\beta_{00}}$ and the $GHZ$-states from 3 to 5 qubits. Figure \ref{fig:ghz_circuit} shows an example of a $GHZ$-state of 3 qubits and its generator circuit. The circuit has a well-defined structure. It starts with a Hadamard gate followed by CNOT gates connecting all subsequent qubits.
    
    These states were chosen because of their relevance to quantum computing and because their circuits are well-known in the literature.

    We performed each simulation with a different number of episodes. The total number of episodes was chosen experimentally, increasing the value until enough successful results were obtained as explained in the next section.

\section{Results and Analysis}
\label{sec:results}

    As said in the previous section, we made 4 simulations in total. Our results are summarized in Table \ref{tab:results}. The first simulation had the 2-qubit state $\ket{\beta_{00}}$ as a goal. After 1000 episodes the agent found 26 different circuits capable of generating $\ket{\beta_{00}}$. The number of circuits found is not equal to the total number of successful circuits that the agent synthesized. Since the agent was able to exploit the same answer for its reward, we counted only the distinguished successful circuits.
    
    \begin{table*}
        \centering
        \begin{tabular}{ccc}
                \toprule
                \textbf{Goal State} & \textbf{Circuits Found} & \textbf{Number of Episodes} \\
                \midrule
                $\ket{\beta_{00}}$ & 26 & 1000\\
                GHZ 3 qubits & 31 & 5000\\
                GHZ 4 qubits & 7 & 20000\\
                GHZ 5 qubits & 1 & 30000\\
            \bottomrule
        \end{tabular}
        \caption{Results obtained by the Projective Simulation Agent. For the goal state $\beta_{00}$ composed of two qubits, the agent was able to find 26 distinct circuits after 1000 episodes. For the 3 qubit GHZ state, the agent was able to found 31 circuits after 5000 episodes. The circuits obtained dropped significantly for the GHZ states with 4 and 5 qubits, only 7 and 1 distinct circuits were find respectively, even after longer runs.}
        \label{tab:results}
    \end{table*}
    
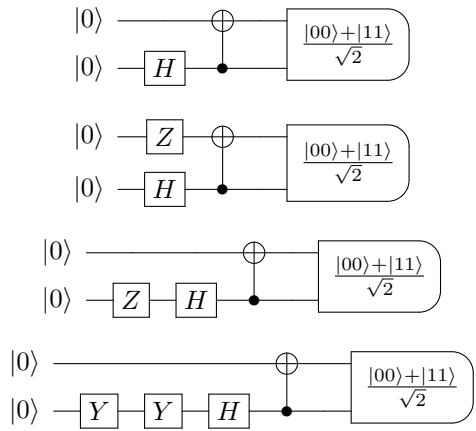
\begin{figure}[h!]
        \centerline{
            \Qcircuit @C=1em @R=.7em {
                \lstick{\ket{0}} & \qw      & \targ     & \qw & \multimeasureD{1}{\frac{\ket{00}+\ket{11}}{\sqrt{2}}} \\
                \lstick{\ket{0}} & \gate{H} & \ctrl{-1} & \qw & \ghost{\frac{\ket{00}+\ket{11}}{\sqrt{2}}}
            }
        }
        
        \bigskip \centerline{
            \Qcircuit @C=1em @R=.7em {
                \lstick{\ket{0}} & \gate{Z} & \targ & \qw  & \multimeasureD{1}{\frac{\ket{00}+\ket{11}}{\sqrt{2}}} \\
                \lstick{\ket{0}} & \gate{H} & \ctrl{-1} & \qw  & \ghost{\frac{\ket{00}+\ket{11}}{\sqrt{2}}}} }
    
        \bigskip \centerline{
            \Qcircuit @C=1em @R=.7em {
                \lstick{\ket{0}} & \qw & \qw & \targ  & \qw   & \multimeasureD{1}{\frac{\ket{00}+\ket{11}}{\sqrt{2}}}\\
                \lstick{\ket{0}} & \gate{Z} & \gate{H} &\ctrl{-1} & \qw & \ghost{\frac{\ket{00}+\ket{11}}{\sqrt{2}}}} }
    
        \bigskip \centerline{
            \Qcircuit @C=1em @R=.7em {
                \lstick{\ket{0}} & \qw & \qw  & \qw      & \targ     & \qw   & \multimeasureD{1}{\frac{\ket{00}+\ket{11}}{\sqrt{2}}}\\
                \lstick{\ket{0}} & \gate{Y} & \gate{Y}  & \gate{H} & \ctrl{-1} & \qw & \ghost{\frac{\ket{00}+\ket{11}}{\sqrt{2}}}} }
    
        \caption{2-qubits circuits found by the synthesizer.}
        \label{fig:2q-cf}
    \end{figure}

    Figure \ref{fig:2q-cf} shows a sample of the circuits found by the agent. The circuit at the top follows the same structure as the basic circuit in Figure \ref{fig:circuit-example}, which is the minimal circuit possible. The agent has also found equivalent circuits but they were not as efficient. The Z gate in the middle circuits does not have any effect when applied to a qubit with value $\ket{0}$, and both Y gates at the bottom circuit cancel each other. These redundant gates can be easily spotted by a later optimization operation.

    The other simulations had as a goal the $GHZ$-states from 3 to 5 qubits. The second simulation found 31 circuits in 5000 episodes. The agent was able to find the minimum circuit showed in Figure \ref{fig:circuit-example}. The third simulation found 7 circuits capable of generating the 4-qubit $GHZ$-state in 20000 episodes.
    
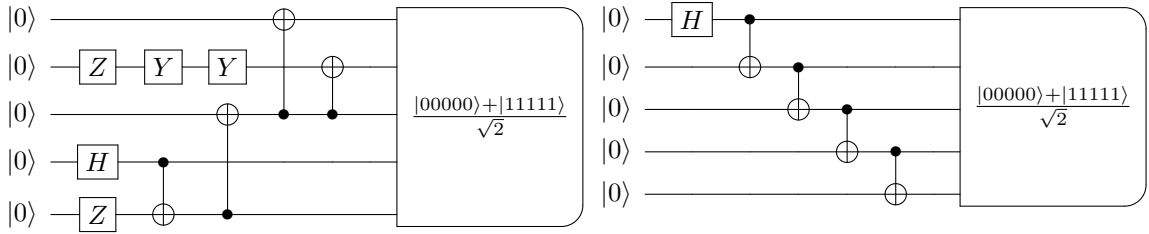
\begin{figure}[h]
        \centerline{
            \Qcircuit @C=1em @R=.7em {
                \lstick{\ket{0}} & \qw      & \qw       & \qw       & \targ     & \qw       & \qw & \multimeasureD{4}{\frac{\ket{00000}+\ket{11111}}{\sqrt{2}}} \\
                \lstick{\ket{0}} & \gate{Z} & \gate{Y}  & \gate{Y}  & \qw       & \targ     & \qw & \ghost{\frac{\ket{00000}+\ket{11111}}{\sqrt{2}}} \\
                \lstick{\ket{0}} & \qw      & \qw       & \targ     & \ctrl{-2} & \ctrl{-1} & \qw & \ghost{\frac{\ket{00000}+\ket{11111}}{\sqrt{2}}} \\
                \lstick{\ket{0}} & \gate{H} & \ctrl{1}  & \qw       & \qw       & \qw       & \qw & \ghost{\frac{\ket{00000}+\ket{11111}}{\sqrt{2}}} \\
                \lstick{\ket{0}} & \gate{Z} & \targ     & \ctrl{-2} & \qw       & \qw       & \qw & \ghost{\frac{\ket{00000}+\ket{11111}}{\sqrt{2}}}
                }
                \hspace{2em}
            \Qcircuit @C=1em @R=.7em
                { \lstick{\ket{0}} & \gate{H} & \ctrl{1} & \qw      & \qw      & \qw      & \qw & \multimeasureD{4}{\frac{\ket{00000}+\ket{11111}}{\sqrt{2}}} \\
                \lstick{\ket{0}}   & \qw      & \targ    & \ctrl{1} & \qw      & \qw      & \qw & \ghost{\frac{\ket{00000}+\ket{11111}}{\sqrt{2}}} \\
                \lstick{\ket{0}}   & \qw      & \qw      & \targ    & \ctrl{1} & \qw      & \qw & \ghost{\frac{\ket{00000}+\ket{11111}}{\sqrt{2}}} \\
                \lstick{\ket{0}}   & \qw      & \qw      & \qw      & \targ    & \ctrl{1} & \qw & \ghost{\frac{\ket{00000}+\ket{11111}}{\sqrt{2}}} \\
                \lstick{\ket{0}}   & \qw      & \qw      & \qw      & \qw      & \targ    & \qw & \ghost{\frac{\ket{00000}+\ket{11111}}{\sqrt{2}}}
                }    
                }
        \caption{5-qubit circuit found by the synthesizer (left). Basic circuit for the 5-qubit GHZ-state (right)}
        \label{fig:5q-cf}
        
        \end{figure}

    The fourth simulation found only one circuit in 30000 episodes. The 5-qubit circuit showed in Figure \ref{fig:5q-cf}, demonstrates that the agent was able to adapt well to the environment's architecture. Depending on the architecture, some qubits may not be directed connected or have only a one-way connection. This fact limits the possible placements for controlled-gates. Since the example circuit was not allowed by the architecture, the agent had to find another sequence of CNOT operations that could achieve the goal state.

\begin{figure*}[h!]
        \centering
        \includegraphics[width=0.45\columnwidth, keepaspectratio]{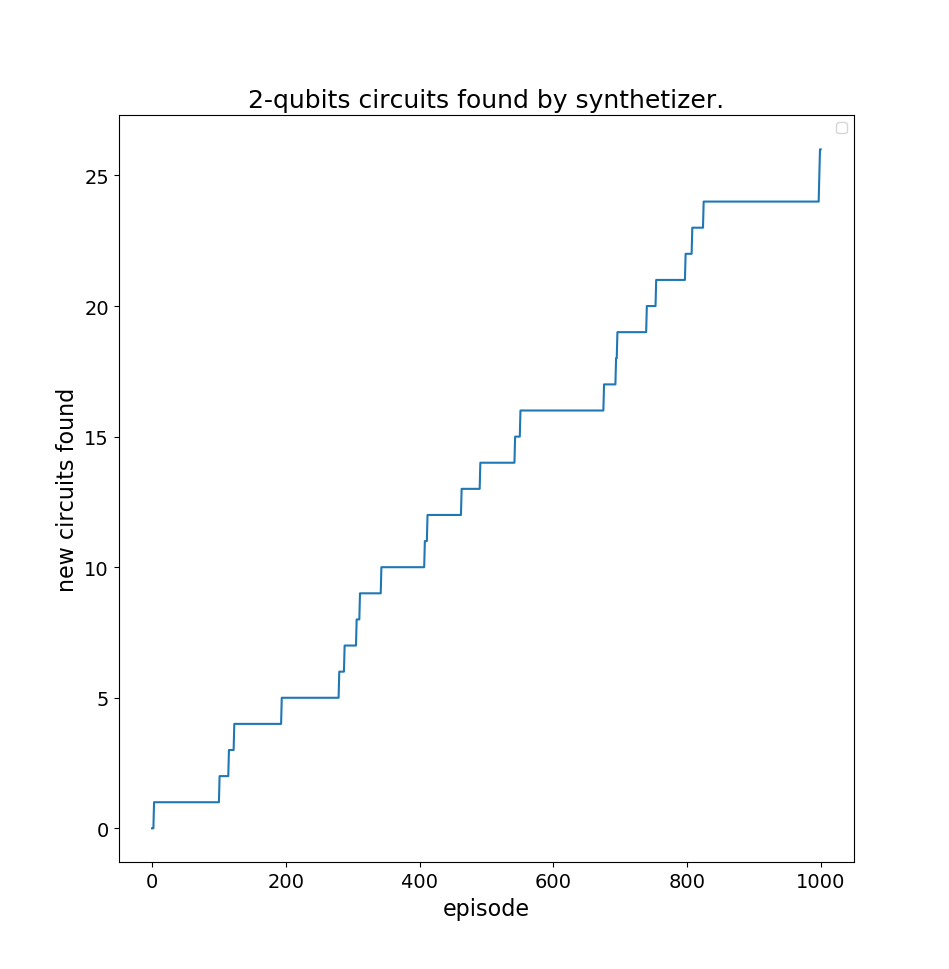}
        \includegraphics[width=0.45\columnwidth, keepaspectratio]{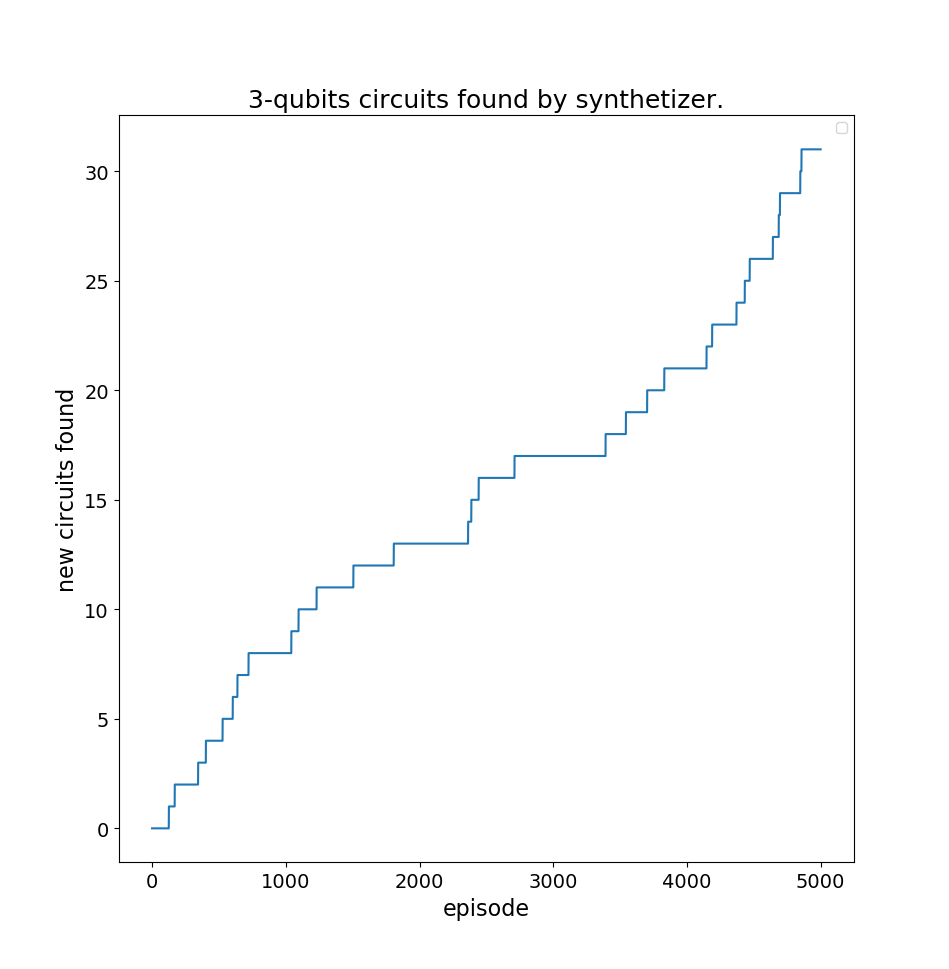}
        
        \includegraphics[width=0.45\columnwidth, keepaspectratio]{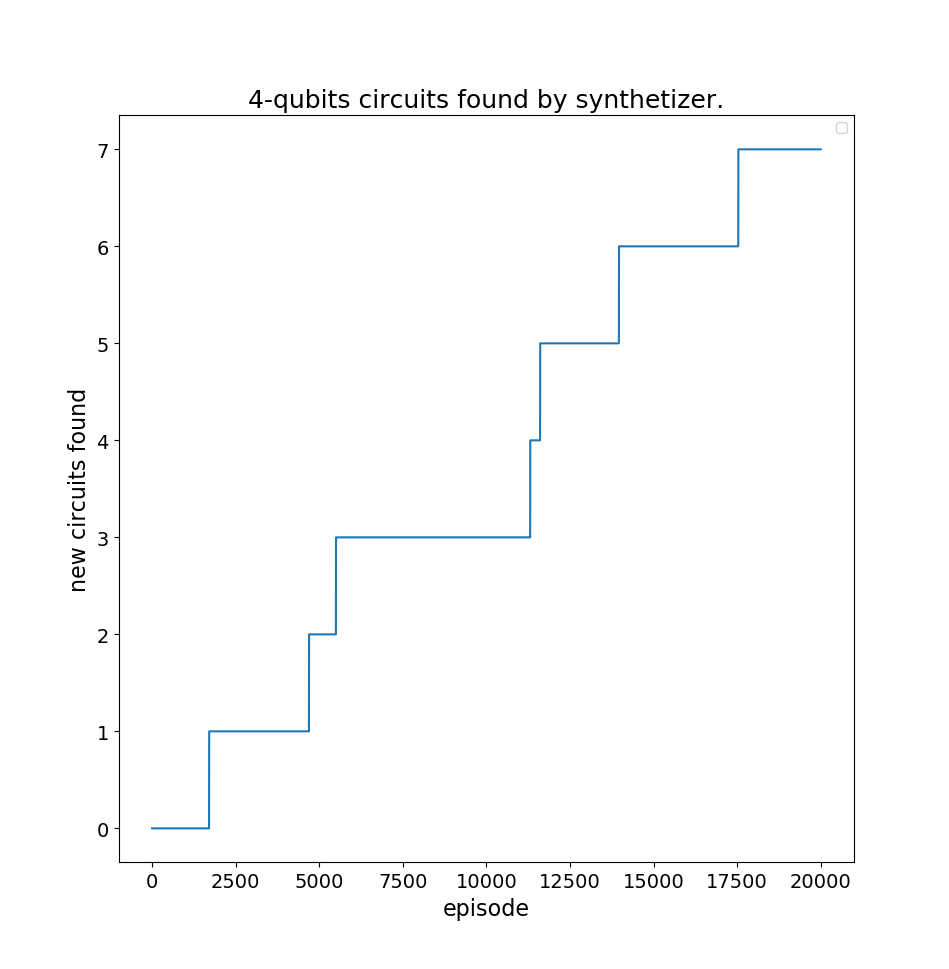}
        \includegraphics[width=0.45\columnwidth, keepaspectratio]{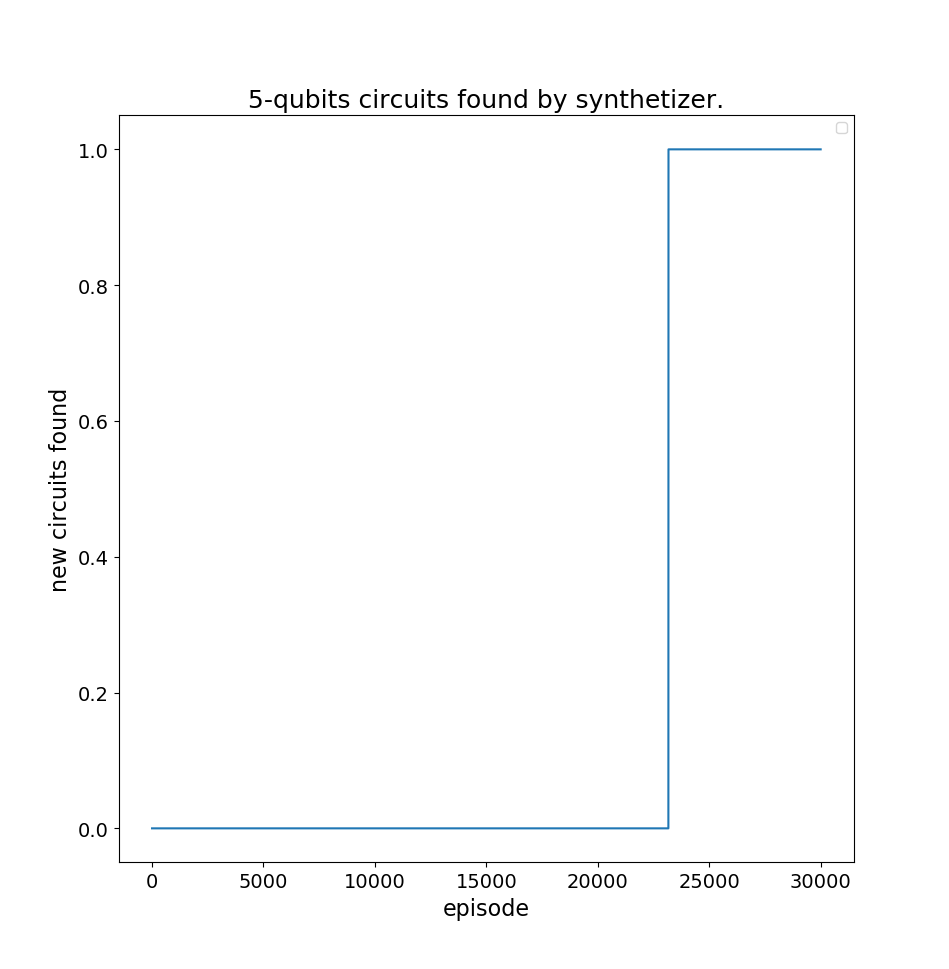}
        
        \caption{Circuits found by synthesizer.}
        \label{fig:graph-cf}
    \end{figure*}

  The learning process of the agent required it to first test possible solutions randomly, and as it finds correct answers it starts to learn the right structure of the circuit. As shown by the graphs in Figure \ref{fig:graph-cf}, the agent had good results for 2-qubits circuits and 3-qubits circuits, but as the number of qubits increased it took more time for the agent to find a correct solution to work with. In the gaps of the graphs, where the agent has not found new circuits, the damping parameter was still active. This fact made the reinforcement that the edges got from the successful circuits fade, so that the learning process almost returned to the initial random search. Such behavior is intensified as the search space increase.
    
    A possible way to overcome the lack of results could be to start the agent with an ECM that already has some information learned to prevent it to search blindly for a first successful circuit. For example, for circuits with a well-defined structure, i.e. W states or GHZ states, the trained ECM of the agents for circuits with less qubits could be loaded and altered with the necessary additional clips. This way, the new agent could start searching by using the previous structure already learned by its predecessor.
    Also, a deeper analysis could show that a lower damping parameter $\gamma$ and a higher reward would help the agent to learn faster after a few successful results are found. The ideal values for the parameters $\gamma$ and $\eta$ can be learned through the process of meta-learning, as showed in \cite{meta-PS}.   

\section{Conclusions}
\label{sec:conclusion}

    We proposed the use of reinforcement learning to approach quantum circuit synthesis. We used projective simulation as the model of our work because it presented good results in discovering new quantum experiments, as shown in \cite{Zeilinger:2018}. 
    At this moment, only a subset of functions from projective simulation was implemented, once we intend to verify the usability of this technique in quantum circuit synthesis.
    
    To ascertain the viability of our model we ran 4 simulations where our agent had to learn how to generate an entangled state with a different number of qubits. As intended, the model learned to build a quantum circuit that was equivalent to the circuit known in the literature. However, the agent found less solutions as the number of qubits increased. We propose two ways to overcome this problem: the use of better parameters $\gamma$ and $\eta$ and to start the agent with an ECM that already has some information learned.
    
    The agent also found alternatives circuits that were different from the basic circuit known in the literature. However, due to the agent's limited action-pool and the simplicity of our model, the alternatives were not efficient, because they had redundant gates. Still, the alternative circuits could be easily minimized by a later optimization process.
    
    For future works, there are more functions available in projective simulation that could be implemented in our agent. For example, the action-composition function could help with more complex circuits, as it allows to find subroutines and to create action clips that serve as a shortcut to them.
    
    Our agent can also be used to optimize circuits. Since only the perceptions that resulted in a rewarded circuit were added definitely in the agent's memory, it is possible to analyze the strongest edges of the memory to find an optimal circuit. As all the circuits that reached the agent's goal may share a sequence of gates that creates the right answer, this sequence would be rewarded in all circuits. As this sequence is likely to not have any unnecessary gates, finding it may optimize the results found by the agent.
    
    We expect reinforcement learning become an important tool in quantum circuit synthesis. The ability to learn new circuit layouts demonstrated that it can be successfully used in order to synthesize more efficient circuits, as seen in \cite{GHZ-W-synth}. Besides that, projective simulation presented here considers the limitations of qubit connectivity when designing a circuit and the quality of quantum gates, which is a factor to still be considered in current research \cite{Mosca_2020}.
    

\section*{Acknowledgements}
EID acknowledges the financial support by Brazilian agencies CAPES, CNPq, and INCT-IQ (National Institute of Science and Technology for Quantum Information).

\bibliographystyle{plain}
\bibliography{biblio}         

\end{document}